\def\wk{\omega_k}
\def\br{{\bf r}}
\def\bR{{\bf R}}
\def\bD{{\bf D}}
\def\brp{{\bf r'}}
\def\bk{{\bf k}}
\def\bkp{{\bk '}}
\def\bmu{\mbox{\boldmath $\mu$}}

\documentclass[preprint,aps,showpacs]{revtex4}

\begin{document}

\title{Field fluctuations near a conducting plate and Casimir-Polder forces
in the presence of boundary conditions}

\author{S. Spagnolo, R. Passante and L. Rizzuto}

\affiliation{CNISM and Dipartimento di Scienze Fisiche ed
Astronomiche, \\
Universit\`{a} degli Studi di Palermo, \\
Via Archirafi 36, I-90123 Palermo, Italy}

\email{roberto.passante@fisica.unipa.it}


\begin{abstract}
We consider vacuum fluctuations of the quantum electromagnetic
field in the presence of an infinite and perfectly conducting
plate. We evaluate how the change of vacuum fluctuations due to
the plate modifies the Casimir-Polder potential between two atoms
placed near the plate. We use two different methods to evaluate
the Casimir-Polder potential in the presence of the plate. They
also give new insights on the role of boundary conditions in the
Casimir-Polder interatomic potential, as well as indications for
possible generalizations to more complicated boundary conditions.
\end{abstract}

\maketitle

\section{Introduction}
\label{sec:1}

Electromagnetic vacuum fluctuations have remarkable effects on
atomic or molecular systems, such as energy level shifts and
spontaneous decay. Another important consequence of the existence
of field fluctuations in the vacuum are Casimir-Polder forces,
which are long-range interactions between neutral atoms or
molecules \cite{CP48}. Casimir-Polder forces are also related to
the Casimir effect, that recently has been measured with
remarkable precision \cite{BMM01,LR02}. The Casimir-van der Waals
interaction between an atom and a dielectric wall has also been
measured \cite{DD03,AD02}.

It is well known that vacuum fluctuations change when boundary
conditions are present, due to the change of the field modes
\cite{KL91,Milton01,WG97}. The aim of this paper is to study how
the physical properties of the vacuum state, in particular the
fluctuations of the electric field, are modified by the presence
of an infinite perfectly conducting plate, and how these changes
reflect upon the Casimir-Polder interaction between two atoms
placed near the conducting plate. The methods used in this paper
give also useful indications on possible extensions to more
complicated cases, such as different geometries of the boundary
conditions or dielectric rather than metallic objects. The results
obtained in this paper also provide new physical models which may
be helpful for understanding the origin and peculiarities of
Casimir-Polder forces with boundary conditions, such as the
suggested enhancement of these forces between atoms inside a
dielectric slab \cite{MD05}.

\section{Field fluctuations near a perfectly conducting plate}
\label{sec:2}

In the multipolar coupling scheme, the Hamiltonian describing two
atoms $A$ and $B$ interacting with the radiation field in the
dipole approximation is
\begin{eqnarray}
H &=& H_F + H_A + H_B + H_{AF} +
H_{BF} \nonumber \\
&=& \sum_{\mathbf{k}j} \hbar \wk a^{\dagger}_{\bk j} a_{\bk j} +
H_A + H_B-\bmu_A \cdot \bD (\br_A)- \bmu_B \cdot \bD (\br_B)
\label{eq:1}
\end{eqnarray}
where
\begin{equation}
\bD(r) = \sum_{\bk j} \bD (\bk j,\br) = i\sum_{\bk j}\sqrt{\frac{2
\pi \hbar \wk}{V}} {\bf f}(\bk j; \br) \left( a_{\bk j} -
a^{\dagger}_{\bk j}\right) \label{eq:2}
\end{equation}
is the transverse displacement field operator, that in this
coupling scheme is the momentum conjugate to the vector potential,
and ${\bf f}(\bk j; \br)$ are the appropriate mode functions
taking into account the boundary conditions for the field
operators (j is a polarization index) \cite{PT82}. The mode
functions, which we take as real functions, satisfy the
normalization condition
\begin{equation}
\frac 1V \int d^3 r {\bf f}(\bk  j; \br){\bf f}(\bkp  j'; \br)=
\delta_{\bk \bkp} \delta_{jj'} \label{eq:3}
\end{equation}
Compared to the free-space case, the presence of the conducting
plate yields a change of vacuum fluctuations, and consequently
changes of radiative processes of atoms placed near the plate
(e.g. level shifts and decay rates) \cite{KL91}; another
consequence is a Casimir-like potential energy between an atom and
the plate \cite{Barton88}. Due to the boundary conditions, the
energy density of vacuum fluctuations also changes
\cite{WG97,SF05}, and this is of course related to the atom-wall
Casimir potential. We shall show that another direct consequence
is a change in the Casimir-Polder intermolecular potential between
two atoms compared to the free-space case.

Vacuum fluctuations have equal-time spatial field correlations,
whose properties are related to the Casimir-Polder potential
between two neutral atoms \cite{PT93}. Field correlations are also
affected by the boundary conditions, as in the present case of a
perfectly conducting plate. This yields a modification of the
Casimir-Polder intermolecular potential between two atoms, as we
shall discuss in detail in the next Section. In the vacuum state
$\mid \{ 0_{\bk j} \} \rangle$, the equal-time spatial correlation
of the displacement field at points $\br_A$ and $\br_B$ is given
by
\begin{equation}
\langle \{ 0_{\bk j} \} \mid D_\ell (\bk j; \br_A) D_m (\bk j;
\br_B) \mid \{ 0_{\bk j} \} \rangle = \frac {2\pi \hbar ck}V
\sum_{\bk j} f_\ell (\bk
 j; \br_A) f_m (\bk  j; \br_B) \label{eq:4}
\end{equation}
In the case of an infinite metallic plate, the sum over
polarizations and the angular part of the integration over $\bk$
can be easily performed, obtaining \cite{PT82}
\begin{equation}
\frac 1{4\pi}\int d\Omega_k \sum_j f_\ell(\bk j;\br_A)f_m(\bk
j;\br_B)=\tau_{\ell m}(kR)-\sigma_{\ell n} \tau_{nm} (k\bar{R})
\label{eq:5}
\end{equation}
where we have introduced the tensor
\begin{eqnarray}
\tau_{\ell m}(k R)&=&\frac 1{4\pi}\int d\Omega_k\left(
\delta_{\ell m}-\hat{k}_\ell \hat{k}_m \right)e^{\pm
i\mathbf{k}\cdot \mathbf{R}}
\nonumber \\
&=& \left(-\nabla^{2}\delta_{\ell m}+\nabla_\ell \nabla_m
\right)^R \frac {\sin k R}{k^3R} \label{eq:6}
\end{eqnarray}
with $R= \mid \brp - \br \mid$ and $\bar{R} = \mid \brp -\sigma
\br \mid$. The matrix
\begin{equation}
\sigma = \left(
\begin{array}{ccc}
1 & 0 & 0 \\
0 & 1 & 0 \\
0 & 0 & -1 \\
\end{array}
\right) \label{eq:7}
\end{equation}
gives a reflection on the conducting plate, supposed orthogonal to
the $z$ axis. The second term of (\ref{eq:5}) is the effect of the
infinite conducting plate on the field correlation function, which
is thus modified with respect to the free-space case. Thus we
expect a modification of the Casimir-Polder potential between two
neutral atoms/molecules when the plate is present.

\section{The Casimir-Polder potential between two atoms
in the presence of a conducting plate}
\label{sec:3}

We now consider the Casimir-Polder long-range interaction between
the two atoms A and B, in the presence of the infinite perfectly
conducting plate. We shall use two different approaches to
calculate the potential, which also give new physical insights on
the origin of the modification of the potential due to the plate.
They also suggest possible extensions to more complicated boundary
conditions. The methods we shall use have already been used for
two- and three-body stationary Casimir-Polder potential
\cite{PPT98,PT93,CP97}, as well as for dynamical Casimir-Polder
forces \cite{RPP04,PPR05}, in the free space.

Using an appropriate transformation, the interaction terms in
Hamiltonian (\ref{eq:1}) can be transformed to the following
effective interaction Hamiltonian \cite{PPT98}

\begin{equation}
H = - \frac 12 \sum_{\bk j}\sum_{\bkp j'} \alpha_A(k)\bD (\bk j,
\br_A) \cdot \bD (\bkp j', \br_A) - \frac 12 \sum_{\bk
j}\sum_{\bkp j'} \alpha_B(k)\bD (\bk j, \br_B) \cdot \bD (\bkp j',
\br_B) \label{eq:8}
\end{equation}
where
\begin{equation}
\alpha(k) = \frac 2{3\hbar c}
\sum_p\frac{k_{p0}\mu^2_{p0}}{k^2_{p0}-k^2}
\label{eq:9}
\end{equation}
is the atomic dynamical isotropic polarizability (for simplicity,
we are assuming isotropic atoms) , $\hbar ck_{p0}=(E_{p}-E_{0})$
is transition energy from atomic state $p$ to the ground state $0$
and $\mu_{p0}$ are matrix elements of the atomic dipole momentum
operator.

In order to calculate the Casimir-Polder potential between the two
atoms in the presence of the conducting plate, we first obtain the
dressed ground state of one atom $(A)$; then we evaluate the
interaction energy of the other atom $(B)$ with the field
fluctations dressing atom $A$. Thus, the first step is to
calculate the dressed ground state of one of the two atoms (let
say A, with position $\br_A$), in the presence of the plate, and
then to evaluate the average value on this state of the effective
interaction Hamiltonian of the second atom (B, with position
$\br_B$) with the field.

The dressed ground state of atom $A$ at the lowest significant
order in the atom-radiation coupling, using straightforward
perturbation theory, is
\begin{equation}
|\{0_{\bk j}\},\downarrow_A \rangle_D= |\{0_{\bk j}
\},\downarrow_A \rangle-\frac{\pi}{V} \sum_{\bk j}\sum_{\bkp
j'}\alpha_{A}(k) \frac{(kk')^{1/2}}{k+k'} \mathbf{f}(\bk
j;\br_A)\cdot \mathbf{f}(\bkp j';\br_A)| 1_{\bk j}1_{\bkp
j'},\downarrow_A \rangle \label{eq:10}
\end{equation}
The interaction energy with atom $B$ is given by the average value
of the effective interaction Hamiltonian of atom $B$ with the
field, evaluated on the dressed ground state of atom A, given by
(\ref{eq:10}),
\begin{eqnarray}
\Delta E_{AB}&=& -\frac{1}{2}\sum_{\bk j} \sum_{\bkp
j'}\alpha_{B}(k)\ _{D}\langle\{0_{\bk j}\}, \downarrow_A |\bD (\bk
j, \br_B)\cdot\bD(\bkp j', \br_B)
| \{0_{\bk j}\},\downarrow_A \rangle_D \nonumber \\
&=& -\frac{2\pi^2\hbar
c}{V^2}\sum_{\bk,\bkp}(\alpha_A(k)+\alpha_A(k'))
\alpha_B(k)\frac{kk'}{k+k'}
\nonumber \\
&\times& \left[ \sum_jf_\ell(\bk j;\br_B)f_m(\bk j;\br_A)\right]
\left[\sum_{j'}f_\ell (\bkp j'; \br_B)f_m(\bkp j';\br_A)\right]
\label{eq:11}
\end{eqnarray}
Using
\begin{equation}
\sum_jf_\ell (\bk j;\br )f_m(\bk j;\br ') =[\delta_{\ell
m}-\hat{k}_\ell \hat{k}_m] e^{i\bk\cdot(\br -\br ')}- \sigma_{\ell
n}[\delta_{nm}-\hat{k}_n \hat{k}_m] e^{i\bk\cdot(\br -\sigma\br
')} \label{eq:12}
\end{equation}
where $\sigma$ is reflection matrix with respect to the plate
(\ref{eq:7}), after substitution of (\ref{eq:12}) into
(\ref{eq:11}) and in the continuous limit, we finally obtain
\begin{eqnarray}
\Delta E_{AB} &=& -\frac {\hbar c}{2(2\pi )^4} \int \!\int d^3k
d^3k'(\alpha_A(k)+\alpha_A(k')) \alpha_B(k)\frac{kk'}{k+k'}
\nonumber \\
&\times& \left( [\delta_{\ell m}-\hat{k}_\ell \hat{k}_m] e^{i\bk
\cdot(\br_B-\br_A)} - \sigma_{\ell n}[\delta_{mn} - \hat{k}_m
\hat{k}_n] e^{i\bk \cdot (\br_B - \sigma \br_A)}\right)
\nonumber \\
&\times& \left( [\delta_{\ell m}-\hat{k'}_\ell \hat{k'}_m]
e^{i\bkp \cdot(\br_B-\br_A)} - \sigma_{\ell p}[\delta_{mp} -
\hat{k'}_m \hat{k'}_p] e^{i\bkp \cdot (\br_B - \sigma
\br_A)}\right) \label{eq:13}
\end{eqnarray}
This is the general expression of the Casimir-Polder potential
energy between the two atoms in the presence of the conducting
plate. In the so-called far zone, that is for interatomic
distances larger than the significant atomic transition
wavelengths from the ground state, we can approximate dynamical
with static polarizabilities $(\alpha_{A,B}(k)\simeq
\alpha_{A,B}(0))$, obtaining
\begin{eqnarray}
\Delta E_{AB} &=& -\frac{\hbar c}{(2\pi )^4}
\alpha_A(0)\alpha_B(0)\int \! \int d^3k d^3k' \frac {kk'}{k+k'}
\nonumber \\
&\times& \left( [\delta_{\ell m}-\hat{k}_\ell \hat{k}_m] e^{i\bk
\cdot \bR} - \sigma_{ln}[\delta_{mn} - \hat{k}_m \hat{k}_n]
e^{i\bk \cdot \bar{\bR}} \right)
\nonumber \\
&\times& \left( [\delta_{\ell m}-\hat{k'}_\ell \hat{k'}_m]
e^{i\bkp \cdot\bR } - \sigma_{\ell p}[\delta_{mp} - \hat{k'}_m
\hat{k'}_p] e^{i\bkp \cdot \bar{\bR}}\right)
\label{eq:14}
\end{eqnarray}
where we have used $\bR = \br_B - \br_A$ and $\bar{\bR} = \br_B -
\sigma \br_A$; $\bar{\bR}$ is the distance between one atom and
the image of the other atom with respect of the plate.

After algebraic calculations we obtain
\begin{eqnarray}
\Delta E_{AB}(R,\bar{R}) &=& -\frac{23}{4\pi}\hbar
c\frac{\alpha_A(0)\alpha_B(0)}{R^7}-\frac{23}{4\pi}\hbar
c\frac{\alpha_A(0)\alpha_B(0)}{\bar{R}^7}
\nonumber \\
&+& \frac{8}{\pi}\hbar
c\frac{\alpha_A(0)\alpha_B(0)}{R^3\bar{R}^3(R+\bar{R})^5} \left(
R^4\sin^2\vartheta+5R^3\bar{R}\sin^2\vartheta \right.
\nonumber \\
&+& \left. R^2\bar{R}^2(6+\sin^2\vartheta+\sin^2\bar{\vartheta})+
5R\bar{R}^3\sin^2\bar{\vartheta}+\bar{R}^4\sin^2\bar{\vartheta}\right)
\label{eq:15}
\end{eqnarray}
where $\vartheta$ and $\bar{\vartheta}$ are respectively the
angles of $\bR$ and $\bar{\bR}$ with the normal to the plate. We
note the presence of three terms: the usual $R^{-7}$
Casimir-Polder potential between the two atoms (as in absence of
the plate), the $\bar{R}^{-7}$ Casimir-Polder-like interaction
between one atom with the reflected image of the second atom, and
a term depending from both variables $R$ and $\bar{R}$. This
result has been already obtained by fourth order perturbation
theory \cite{PT82}. We have now obtained the same result with a
different and simpler method, which stresses the role of dressed
field fluctuations modified by the boundary conditions given by
the presence of the conducting plate.

The potential (\ref{eq:15}) can be also obtained with a different
method, based on the properties of the spatial correlations of
vacuum fluctuations, which are modified by the conducting plate as
discussed in the previous Section. The basic idea is that the
vacuum fluctuations of the electromagnetic field induce
instantaneous dipole moments on the two atoms, and that these
induced dipoles are correlated because vacuum fluctuations are
spatially correlated. The Casimir-Polder potential energy then
arises from the (classical) interaction between the two correlated
induced dipoles \cite{PT93} . The arguments used in this physical
model are essentially classical, except for the assumption of the
real existence of vacuum fluctuations, which invokes quantum
aspects of the electromagnetic field. This makes the model very
interesting from a fundamental point of view, giving insights on
the origin of Casimir-Polder forces and stressing the role of
zero-point fluctuations. In our case, as discussed in the previous
Section, the conducting plate changes the spatial correlations of
vacuum fluctuations and a change of the Casimir-Polder potential
is thus expected. We show that this physical model is indeed
correct, but, compared to the free-space case, a modification of
the classical interaction energy between the two dipoles is also
necessary, in order to take into account the image atoms.

The relation between the induced dipole moments in the atoms,
assumed isotropic for simplicity, and the field is
\begin{equation}
\mu_\ell (\bk ,j)=\alpha(k) D_\ell (\bk j,\br) \label{eq:16}
\end{equation}
The average interaction energy between the induced and correlated
atomic dipoles is
\begin{eqnarray}
V_{AB}&=& \sum_{\bk j}\mu_\ell^A (\bk ,j)\mu_m^B (\bk ,j) V_{\ell
m}(k,R,\bar{R}) \nonumber \\
&=& \sum_{\bk
j}\alpha_A(k)\alpha_B(k) \langle D_\ell (\bk
 j, \br_A) D_m (\bk j,\br_B) \rangle V_{\ell m}(k,R,\bar{R})
\label{eq:17}
\end{eqnarray}
where the average is taken on the ground state of the field; the
presence of the spatial correlation function of the transverse
displacement field should be noted. $V_{lm}(k,R,\bar{R})$ is a
classical interaction energy between the induced dipole moments of
the two atoms. In the case of atoms in the free space, this
interaction energy is the classical interaction energy between
dipoles oscillating at frequency $ck$ \cite{PT93}. When boundary
conditions are present, this potential must be appropriately
changed in order to take into account the interaction with the
image atoms. We therefore take the following form of the classical
interaction energy
\begin{eqnarray}
V_{\ell m}(k,R,\bar{R})&=& V_{\ell m}(k,R)-\sigma_{\ell
p}V_{pm}(k,\bar{R})
\nonumber \\
&=&(\nabla^{2}\delta_{\ell m}-\nabla_\ell\nabla_{m})^{R}\frac{\cos
kR}{R}-\sigma_{\ell p}(\nabla^{2}\delta_{pm}-\nabla_{p}
\nabla_{m})^{\bar{R}}\frac{\cos k\bar{R}}{\bar{R}} \label{eq:18}
\end{eqnarray}
where the superscript of the differential operators above indicate
the variable on which the operators act. The first terms is the
same used for atoms in the free space. The second term arises from
the presence of the conducting plate, and has the form of a
atom-image interaction.

Using (\ref{eq:4}) the interaction energy (\ref{eq:17}) becomes
\begin{equation}
V_{AB} = \sum_\bk \alpha_A(k)\alpha_B(k)\frac {2\pi \hbar ck}V
\left( \sum_j f_\ell (\bk j; \br_A) f_m (\bk j; \br_B)\right)
V_{\ell m}(k,R,\bar{R})
\label{eq:19}
\end{equation}
In the continuous limit
\begin{equation}
\sum_{\bk} \Longrightarrow \frac{V}{(2\pi)^3}\int d\Omega_k\int
k^2dk
\label{eq:20}
\end{equation}
and using (\ref{eq:5},\ref{eq:6}), we obtain
\begin{equation}
V_{AB}=\frac{\hbar c}\pi\int_0^{\infty} \!
dk{\,}k^3\alpha_A(k)\alpha_B(k) \left( \tau_{\ell
m}(kR)-\sigma_{\ell n}\tau_{nm}(k\bar{R})\right) \left( V_{\ell
m}(k,R)-\sigma_{\ell p}V_{pm}(k,\bar{R})\right) \label{eq:23}
\end{equation}
Taking into account the familiar expression of the Casimir-Polder
interaction between two atoms in the free space \cite{CPP95}
\begin{eqnarray}
V_{CP}(R)&=& \frac{\hbar c}{\pi}\int_0^{\infty}dk{\,}k^3
\alpha_A(k)\alpha_B(k) \tau_{\ell m}(kR)V_{\ell m}(k,R)
\nonumber \\
&=& -\frac{\hbar c}{\pi R^2}\int_0^{+\infty} \!
du{\,}u^4\alpha_A(iu)\alpha_B(iu)\left(1+\frac{2}{uR}+\frac{5}{u^2R^2}+
\frac{6}{u^3R^3}+\frac{3}{u^4R^4}\right)e^{-2uR}
\label{eq:24}
\end{eqnarray}
and also that $\sigma_{\ell n} \tau_{nm}(k\bar{R}) \sigma_{\ell p}
V_{pm}(k,\bar{R}) =\tau_{\ell m}(k\bar{R)}V_{\ell m}(k,\bar{R)})
$, we immediately see from (\ref{eq:23}) the presence of
$V_{CP}(R)$ and $V_{CP}(\bar{R})$, plus two extra terms with both
variables $R$ and $\bar{R}$. $V_{CP}(R)$ and $V_{CP}(\bar{R})$ are
respectively the interaction between the two atoms, as in the
absence of the conducting plate, and the interaction between one
atom and the image of the other atom. The other terms in
(\ref{eq:23}) contain both coordinates $R$ and $\bar{R}$, and
their physical interpretation is not so evident. When
(\ref{eq:23}) is approximated to the far zone, replacing the
dynamical polarizabilities with the static ones, the same result
of equation (\ref{eq:15}) is obtained.

This result shows how the method for the calculation of
Casimir-Polder forces based on field correlations must be modified
when boundary conditions are present. In fact, there are two
elements that must be considered. First, the expectation value of
the field correlation function changes in the presence of the
boundary condition compared to the case of the unbounded space, as
expressed by (\ref{eq:4}). Secondly, the classical interaction
energy between the correlated induced dipoles must be changed
according to (\ref{eq:18}), in order to take into account the
image dipoles too. This also gives helpful indications for a
generalization to more complicated boundary conditions, such as
cavities or dielectric objects.

The two methods we have used to calculate the Casimir-Polder
potential yield the same result. Both methods are based on the
common idea that electromagnetic field fluctuations induce real
effects in matter. The mathematical equivalence of the two methods
can be proved formally, similarly to the case of two atoms in the
free space \cite{PPR03}. In fact, after some algebraic
manipulation eq. (\ref{eq:11}) can be expressed in the following
form
\begin{eqnarray}
\Delta E_{AB}&=& -\frac{4\pi^2\hbar c}{V}\sum_{\bk j}k f_\ell (\bk
j;\br_B)f_m(\bk j;\br_A) \nonumber \\
&\times& \left[ \alpha_A(k)\frac{1}{V}\sum_{\bkp
j'}\frac{k'^2}{k'^2-k^2} f_\ell (\bkp j';\br_B)f_m(\bkp
j';\br_A)\alpha_B(k')+ \right.
\nonumber\\
&+&\left. \frac{1}{V}\sum_{\bkp j'}\frac{k'^2}{k'^2-k^2} f_\ell
(\bkp j';\br_B)f_m(\bkp j';\br_A)\alpha_A(k')\alpha_B(k')\right]
\label{eq:25}
\end{eqnarray}
In the continuous limit, after evaluation of angular integrals, we
obtain
\begin{eqnarray}
\Delta E_{AB}&=& \frac{2\pi\hbar c}{V}\sum_{\bk j}k f_\ell (\bk
j;\br_B)f_m(\bk j;\br_A)
\nonumber \\
&\times& \alpha_A(k)\alpha_B(k)[(\nabla^{2}\delta_{\ell
m}-\nabla_\ell\nabla_{m})^{R} \frac{\cos kR}{R}-\sigma_{\ell
p}(\nabla^{2}\delta_{pm}-
\nabla_{p}\nabla_{m})^{\bar{R}}\frac{\cos k\bar{R}}{\bar{R}}]
\nonumber\\
&=&\frac{2\pi\hbar c}{V}\sum_{\bk j}k f_\ell (\bk j;\br_B)f_m(\bk
j;\br_A) \alpha_A(k)\alpha_B(k)[V_{\ell m}(k,R)-\sigma_{\ell
p}V_{pm}(k,\bar{R})]
\nonumber\\
&=&\frac{2\pi\hbar c}{V}\sum_{\bk j}\alpha_A(k)\alpha_B(k)\langle
0_{\bk j} \mid D_\ell (\bk j, \br_A) D_m (\bk j, \br_B) \mid
0_{\bk j} \rangle V_{\ell m}(k,R,\bar{R}) \label{eq:26}
\end{eqnarray}
This relation indeed shows that the intuitive model based on
spatial field correlations can be derived from the Hamiltonian
(\ref{eq:1}), even when a boundary conditions such as an infinite
conducting plate is present. This gives further support to this
physical model, and we expect it should be valid also in the case
of more complicated boundary conditions.

\section{Conclusions}
\label{sec:4}

In this paper we have first considered vacuum field fluctuations
in the presence of a perfectly conducting plate, and discussed how
they are modified compared with the free-space case. We have then
considered the modification to the Casimir-Polder intermolecular
potential between two atoms due to the presence of the plate, as a
result of the modified zero-point fluctuations . We have
calculated the Casimir-Polder potential using two different
methods, one based on dressed vacuum fluctuations and the other on
spatial vacuum field correlations. Our results agree with previous
results obtained by fourth-order perturbation theory. Our methods
have however two advantages. The first is that they are
mathematically simpler, and this may be particularly relevant in
more complicated situations; the second is that they give new
physical insights on the role of boundary conditions on
Casimir-Polder forces. Furthermore, the methods used in this paper
give indications on possible generalizations to more complicated
boundary conditions.

\begin{acknowledgments}
The authors wish to thank F. Persico for many discussions and
suggestions on the subject of this paper. This work was in part
supported by the bilateral Italian-Belgian project on
``Casimir-Polder forces, Casimir effect and their fluctuations"
and by the bilateral Italian-Japanese project 15C1 on ``Quantum
Information and Computation" of the Italian Ministry for Foreign
Affairs. Partial support by Ministero dell'Universit\`{a} e della
Ricerca Scientifica e Tecnologica and by Comitato Regionale di
Ricerche Nucleari e di Struttura della Materia is also
acknowledged.
\end{acknowledgments}

\end{document}